\documentclass[aps,prl,twocolumn, showpacs,amsmath,amssymb,amsfonts,nofootinbib,long]{revtex4}
\usepackage{epsfig,latexsym}

\newcommand\BBN{\mbox{\tiny{BBN}}}
\newcommand\QCD{\mbox{\tiny{QCD}}}
\newcommand\GUT{\mbox{\tiny{GUT}}}
\newcommand{\mq}{m_{\QCD}}
\newcommand{\mb}{m_B}
\newcommand{\ov}{\overline}
\newcommand{\bra}{\langle}
\newcommand{\ket}{\rangle}

\newcommand{\Pl}{\mbox{\scriptsize Pl}}
\newcommand{\mc}{\mathcal}

\newcommand\G{\mbox{G}}

\newcommand\kpc{\mbox{kpc}}
\newcommand\pc{\mbox{pc}}
\newcommand\eV{\mbox{eV}}

\newcommand\MeV{\mbox{MeV}}
\newcommand\GeV{\mbox{GeV}}

\newcommand\aem{\alpha_{\rm em}}

\newcommand\f{f_{12}}

\newcommand\T{\Theta}
\newcommand\Tb{\ov \Theta}

\begin{document}

\title{Primordial Magnetic Fields and the Peccei-Quinn Scale}
\author{L. Campanelli$^{1,2}$}
\email{campanelli@fe.infn.it}
\author{M. Giannotti$^{1,2}$}
\email{giannotti@fe.infn.it}

\affiliation{$^{1}${\it Dipartimento di Fisica, Universit\`a di Ferrara, I-44100 Ferrara, Italy
\\           $^{2}$INFN - Sezione di Ferrara, I-44100 Ferrara, Italy}}

\date{November 2006}


\begin{abstract}
A strong primordial magnetic field can induce a relaxation of the
present bound on the PQ-constant. We show that, considering the
present limits on primordial magnetic fields, a value for the
PQ-constant very close to the GUT scale is not excluded. This
result naturally opens the possibility for the axion to be defined
in the context of the GUT theories.
\end{abstract}


\maketitle


After about 30 years, the Peccei-Quinn (PQ) mechanism~\cite{PQ} is
still the most appealing solution of the strong-CP problem (for a
review see, e.g., Ref.~\cite{Kim}), which consists in explaining
the smallness of the CP-violation induced by the QCD Lagrangian.
This violation resides in the presence of the so called $\T-$term,
$\mathcal{L}_{\rm CP} =( \alpha_s/8 \pi)\Tb \, G\widetilde G$,
where $\alpha_s$ is the fine structure constant of the strong
interactions, while $G$ and $\widetilde G$ are the gluon field and
its dual. Experimental limits on the neutron's electric dipole
moment, lead to the unnaturally small upper bound $\Tb \lesssim
10^{-10}$.

The fundamental new feature of the PQ-mechanism is the existence
of an extra global (axial) symmetry, $U(1)_{\rm PQ}$,
spontaneously broken at some energy scale $f_a$, known as the
PQ-scale (or PQ-constant). Therefore, at energies below $f_a$ a
new phase degree of freedom $\T(x)=a(x)/f_a$ emerges as the
Goldstone mode of the $U(1)_{\rm PQ}$ symmetry. This new field
$a$, known as axion, is the most relevant prediction of the
PQ-mechanism. Quantum effects (chiral anomaly), which explicitly
break the PQ symmetry, generate a non trivial axion-gluon
interaction of the form
$\mathcal{L}_{ag} = (\alpha_s/8 \pi) (a/f_a)G \widetilde G$,
and correspondingly an axion potential $V(a)$. The Vafa-Witten
theorem~\cite{Vafa} states that this is minimized when the
Lagrangian is CP-even, that is when the vacuum expectation value
of the axion field cancels the original $\Tb$-term in the QCD
Lagrangian, $\bra \T \ket = -\Tb$.

The interaction with gluons induces a mass $\mq$ for the axion
field, which has the constant value
\begin{equation}
\label{m0} \mq(T) = m \simeq 6.2 \,\eV / (f_a/10^{6} \GeV) \;\;\;
\mbox{for} \;\; T \ll \Lambda \, ,
\end{equation}
at energies (temperatures) below the QCD scale $\Lambda \! \sim \!
200 \,\MeV$, while at higher energies it is suppressed
as~\cite{GPY,Kolb}
\begin{equation}
\label{mT} \mq(T) \simeq 0.1 m \left(\Lambda/T\right)^{3.7} \;\;\;
\mbox{for} \;\; T \gg \Lambda \, .
\end{equation}
Besides gluons, axions interact with fermions in a way inversely
proportional to $f_a$ and with photons through the electromagnetic
anomaly
$\mc L_{a\gamma} = (1/4) \, g_{a\gamma} a F \widetilde F$.
Here, $F$ is the electromagnetic field, $\widetilde F$ its dual,
and $g_{a\gamma} = \aem \xi / (2\pi f_a)$, with $\aem$ the
electromagnetic fine structure constant and $\xi$ an order one,
model dependent constant.
Therefore, the axion phenomenology is characterized by the
PQ-constant, a free parameter of the mechanism.
Today, the combined limits from terrestrial experiments, the
stellar evolution and the supernova neutrino signal exclude all
the values of $f_a$ up to $10^{9} \GeV$ \cite{Raff,BGG}.
On the other hand, cosmological considerations exclude the values
for $f_a$ above $10^{12} \GeV$~\cite{Preskill}.

So, the PQ-scale is not related to any of the relevant scales in
high energy physics, being well above the electroweak scale,
$T_{ew} \simeq 250 \GeV$, but also largely below the scale of the
Grand Unification Theory (GUT), $ T_{\GUT} \sim 10^{15 \div 16}
\GeV$. It is therefore not very plausible that the PQ-mechanism
could be related to the physics at these scales, and this is a
rather unattractive feature of this elegant mechanism. What is the
origin and meaning of this new scale? Is it possible to relax the
bounds on the PQ-constant to a more meaningful scale?
We refer to these questions as the {\it PQ-scale problem}. A
discussion of such a problem might seem premature when there is no
experimental evidence that the PQ-mechanism is effectively
realized in nature. However, the relevance of the axion in
cosmology is universally accepted. Excluding supersymmetric
particles, axions are certainly the best candidate for the dark
matter component of the universe. It is, therefore, a rather
unpleasant result that, for axions to represent the dark matter,
it is necessary to fix the PQ-constant to the nowadays {\it
meaningless} scale $f_a \simeq 10^{12} \GeV$. Understanding the
origin of this {\it new} scale, or trying to relax this toward an
already known scale, is certainly one of the most relevant problem
of axion physics. This explains the number of papers that have
been addressed to it (see, e.g.,
Ref.~\cite{Lazarides,Dvali,Maurizio}).

The aim of this paper is to investigate the possibility of
relaxing the upper bound on the axion-constant through the
interaction of the axion with an intense, primordial magnetic
field~\cite{noi2}.


Before proceeding, we shall analyze the present limits on the
intensity of a cosmological magnetic field at the scales relevant
to our problem. Though the origin of the observed large scale
magnetic fields is still unclear, we cannot exclude the existence
of primordial magnetic fields in the early
universe~\cite{Magn1,Magn2}, as long as their presence does not
invalidate the predictions of the standard cosmological model. In
particular, they must satisfy constraints coming from the Big Bang
Nucleosynthesis (BBN) and the Cosmic Microwave Background (CMB)
radiation, which represent the most stringent limits on cosmic
magnetic fields.

Since in the early universe the conductivity of the primordial
plasma, $\sigma$, is very high, magnetic fields are frozen into
the plasma and evolve adiabatically, $B \propto R^{-2}$, where $R
\propto g_{*S}^{-1/3} T^{-1}$ is the expansion factor, and
$g_{*S}$ counts the total number of effectively massless degrees
of freedom referring to the entropy density of the universe
\cite{Kolb}. It is convenient to parameterize the magnetic field
as
\begin{equation}
\label{Limit1} B = b(T) T^2.
\end{equation}
%
%
%
In the following we shall assume that a uniform magnetic field is
present during the evolution of the axion between the electroweak
scale and a few times the QCD-scale, say $T \simeq 1 \GeV$. Since
a magnetic field correlated on the Hubble scale at $T = 1 \GeV$
can be considered as uniform for $1 \GeV \lesssim T \lesssim
T_{ew}$, we will analyze, to be conservative, the limit on
magnetic fields correlated on the Hubble scale at that time, which
corresponds to a comoving scale $\xi_B \simeq 6 \times 10^{-2}
\pc$.
Unfortunately, no limits coming from the BBN and CMB on these
scales exist. The limit coming from BBN refers to uniform magnetic
fields at that time or, equivalently, correlated on the Hubble
comoving scale $L \simeq 1 \kpc$. The upper bound is given in
Ref.~\cite{Grasso}:
$B(T_{\BBN}, L \simeq 1 \kpc) \lesssim 1 \times 10^{11} \G$,
where $T_{\BBN} = 10^9 K \simeq 0.1 \MeV$. The strongest limit on
small-scale magnetic fields from CMB are given in Ref.~\cite{JKO}.
There, it is deduced the limit
$B(T_0,L=400 \, \pc) \lesssim 3 \times 10^{-8} \G$,
$T_0$ being the actual temperature, on a magnetic field correlated
on a comoving scale $L = 400 \pc$.
In order to convert the above limits in a constraint on magnetic
fields correlated on smaller scales (in particular on the scale
$\xi_B$) we must perform a suitable average over the magnetic
domains. Following the standard procedure (see, e.g.
Ref.~\cite{Magn2} and references therein) we can write
$B(T,L) \equiv B(T,\xi_B)/N^p$,
where $N = L/\xi_B$, with $L$ the comoving scales on which we want
average and $\xi_B$ the comoving magnetic correlation length.
Here, $p=1/2,1,3/2$, depending on the statistical properties of
the tangled magnetic field \cite{Magn2}.
\\
In the two cases referring to BBN and CMB, $N \simeq 2 \times
10^4$ and $N \simeq 6 \times 10^3$, respectively. Therefore, the
BBN and CMB limits translate to
$B_{\rm max}(T_{\BBN},\xi_B) \simeq 1 \times 10^{13} \G$,
and
$B_{\rm max}(T_0,\xi_B) \simeq 2 \times 10^{-6} \G$,
respectively, where we have considered the most conservative case
$p=1/2$.
Evolving adiabatically the above maximum values of the magnetic
field back in time, we get that the limit coming from the CMB
analysis is more stringent with respect to the one from BBN of
about two order of magnitude, and gives
$b_{\rm max}(T) \simeq 1.2 \, g_{*S}^{2/3}(T)$.
In particular, the maximum allowed value at the electroweak scale
is
$B_{\rm max}(T_{ew},\xi_B) \simeq 2 \times 10^{25} \G$.
Essentially, the above constraint corresponds to having a magnetic
field whose energy density, in the radiation era, is less than the
energy density of the universe, $\rho = (\pi^2/30) g_{*}(T) \,
T^4$, where $g_{*}$ counts the total numbers of effectively
massless degrees of freedom referring to the energy density of the
universe. In fact, imposing that $\rho_B = B^2/2 \lesssim \rho$,
we get $b(T) \lesssim 0.8 \, g_{*}^{1/2}(T)$. (Note that in the
range of interest, $1 \GeV \lesssim T \lesssim T_{ew}$, the
quantities $g_{*}$ and $g_{*S}$ are equal~\cite{Kolb}.)

Coming back to the axion's cosmology, there are two
phenomenological aspects that should be considered, {\it i}) an
external magnetic field allows the axion to mix with one photon.
However, this does not lead to a relevant change of the axion's
cosmology \cite{Yanagida,paradox,noi} (see also note 2) and will
not be considered in this paper; {\it ii}) in an external magnetic
field the axion has a contribution $m_B$ to its mass
\cite{Mikheev,mass}, so that $m_{\rm tot}^2 = \mq^2 + \mb^2$. As
we will show, this has at least two possible phenomenological
consequences. First, the magnetic field can induce a breaking of
the PQ-symmetry independent of the QCD dynamics, and this might
spoil the PQ-mechanism. This in principle sets an upper limit on
the intensity of the primordial magnetic field. However, as we
shall see, this is not competitive with the bounds from BBN and
CMB.
Second, the magnetic induced mass can force the cosmological axion
production mechanism to start earlier, therefore changing the
expected axion relic abundance.~\footnote{This last aspect was
considered in Ref.~\cite{noi3}, however without accounting for the
temperature effects.}
We will show that, if the primordial magnetic field is
sufficiently intense, this is indeed the case. A consequence is a
relaxation of the cosmological bound on the PQ-constant. Moreover,
the present limits on the intensity of the magnetic field cannot
exclude a value of the PQ-constant very close to the GUT scale.

In order to show what is stated above, let us consider the
cosmological evolution of the axion field $\T$. Today $\T$ is
settled in the CP-conserving minimum $\T_{\rm today} = \Tb$.
However, just after the PQ symmetry breaking, at temperatures of
order of the PQ-scale, the axion potential is flat and the value
of the phase $\T$ is chosen stochastically. We shall indicate this
as $\T_i$. The misalignment between the two angles $\T_i$ and
$\Tb$ is at the origin of an efficient mechanism for the
generation of axions, known as the {\it axion misalignment
production} \cite{Preskill}. The evolution of $\T$ is described by
the equation of motion \cite{Kolb}
\begin{equation}
\label{1} \ddot{\Theta} + 3H \dot{\Theta} + m_{\QCD}^2 (\Theta -
\Tb) = 0,
\end{equation}
where
$H \simeq 1.66 g_{*}^{1/2} T^2/m_{\Pl}$
is the Hubble parameter with $m_{\Pl}$ the Planck mass.
\\
For high temperatures, the mass term in Eq.~(\ref{1}) is
negligibly small compared to the friction (Hubble) term, so the
axion remains frozen to its initial value $\Theta_i$. However, as
the axion mass becomes dominant over the friction term, $\Theta$
begins to oscillate with the frequency $m_{\QCD}$ and will
approach the CP-conserving limit $\T_{\rm today} \sim \Tb$. During
this period of coherent oscillations, if axions are not
interacting, their number, in a comoving volume, remains constant,
so the axion relic abundance today can be easily calculated as
\begin{equation}
\label{Omega} \Omega_a \simeq 1.6 \, \T^2_i \, g_{*1}^{-1/2}
f_{12} \, (\GeV/T_1),
\end{equation}
where $g_{*1} = g_{*}(T_1)$ and $\f = f_{a}/(10^{12} \GeV)$. Here,
the temperature $T_1$ is such that $m_{\QCD}(T_1) = 3H(T_1)$ and
represents, approximately, the time when the oscillations start.
\footnote{The hypothesis that axions are not interacting during
the time of coherent oscillations, on which this cosmological
argument resides, is well justified even in presence of a very
strong magnetic field. In fact, for the problem at hand, the
axion-photon mixing angle is about $g_{a\gamma}B/(2\sigma) \ll 1$.
The resulting axion relic abundance, even in the most optimistic
case would differ from the standard one only by a factor of about
$10^{-10}$~\cite{noi}.}
If the only contribution to the axion mass were given by the QCD
effects (\ref{mT}), then $T_1 \simeq 0.9 \, \Lambda_{200}^{0.65}
f_{12}^{-0.175} \, \GeV$, where $\Lambda_{200} =
\Lambda/(200\MeV)$. As a consequence, Eq.~(\ref{Omega}) reduces to
$\Omega_{a} \simeq 0.2 \, \Lambda_{200}^{-0.65} \Theta_i^2
\f^{1.175}$.
Assuming $\Theta_i \simeq 1$ \cite{Kolb}, we get $\Omega_{a}
\simeq 0.3$ ($i.e.$ the expected dark matter abundance) for $\f
\simeq 1$. Much larger values of $\f$ would cause too much axion
production and are therefore excluded. This observation leads to
the upper limit on the PQ-constant discussed in the literature,
$f_a \lesssim 10^{12} \GeV$ \cite{Preskill,Kolb}.

However, if a strong external magnetic field is present, the axion
mass has a magnetic contribution, which in the range of interest
for the problem at hand, $1 \GeV \lesssim T \lesssim T_{ew} \simeq
250 \, \GeV$, is~\cite{mass}
\begin{equation}
\label{cosmo6} \mb \simeq g_{a\gamma} B \simeq 7.5 \times 10^{-3}
\xi b \, \Lambda_{200}^2 \, m \, (T/\Lambda)^2,
\end{equation}
where in the last equality we used Eq.~(\ref{Limit1}). In order to
compare the electromagnetic and QCD axion masses, it is useful to
introduce the temperature $T_*$ such that $\mb(T_*) = \mq(T_*)$.
It results
$T_* \simeq 1.6 \, \xi^{-0.18} b^{-0.18} \Lambda_{200}^{-0.35}
\Lambda$.
From the above equation we see that for strong magnetic fields,
say $b \sim 1$, the temperature at which the QCD and
electromagnetic axion masses are equal is about $T_* \simeq few
\times \Lambda$ (see Fig.~1). This means that a strong enough
magnetic field induces a contribution to the axion mass which
would dominate the standard QCD one sufficiently above the QCD
phase transition.


\begin{figure}[t]
\noindent\includegraphics[width=8cm,angle=0]{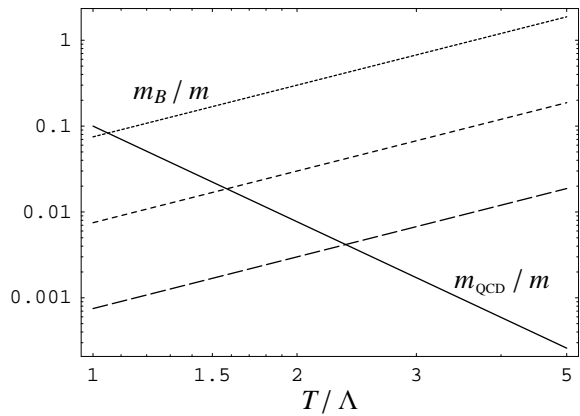}
\caption{Solid line refers to the QCD axion mass, Eq.~(\ref{mT}),
while dotted, dashed, and long-dashed lines refer to the
electromagnetic contribution to the axion mass,
Eq.~(\ref{cosmo6}), for $b=10$, $b=1$, and $b=0.1$, respectively.
Here, we have taken $\xi = \Lambda_{200} =1.$}
\end{figure}


Now, since both $\mb$ and $H$ scale as $T^2$, we can distinguish
two cases. If $\mb < 3H$, that is if
$b < b_{\rm th} \simeq 3.5 \times 10^{-4} \xi^{-1} g_{*}^{1/2} \f$,
the (Hubble) friction is always greater then the electromagnetic
mass. Therefore, the axion coherent oscillations would start only
when the QCD mass equals the friction term. In other words,
$b_{\rm th}$ indicates a threshold value for the parameter $b$,
and therefore for the magnetic field, such that if $b < b_{\rm
th}$ the standard analysis of the axion misalignment production
applies.
On the other hand, for $b > b_{\rm th}$ the electromagnetic mass
term is always greater than the friction term and, consequently,
the axion coherent oscillations start at the time when the
magnetic field is generated.
In the following, to avoid inessential complications, we shall
assume that a magnetic field is generated above or during the
electroweak phase transition. In this case, if $b > b_{\rm th}$,
the axion coherent oscillations would start at $T_{ew}$ (above
that the magnetic-induced axion mass vanishes~\cite{noi3}), and
the present axion relic abundance would be
$\Omega_a \simeq 0.6 \times 10^{-3} \Theta_i^2 f_{12}$.
From the above equation we can deduce the maximum value of the
Peccei-Quinn constant corresponding to the maximum value allowed
for the axion relic abundance, $\Omega_a \simeq 0.3$, which
implies that the dark matter component of the universe is composed
by cold axions. Taking $\Theta_i \simeq \xi \simeq 1$, we see that
this value for the axion abundance is compatible with a
PQ-constant equal to $f_a \simeq 0.5 \times 10^{15} \, \GeV$, a
scale very close to the GUT scale. This value of $f_a$ corresponds
to $b \gtrsim 1.7$, that is $B \gtrsim 1.5 \times 10^{24} \G$ at
the electroweak phase transition. As discussed above, this is
compatible with all constraints coming from cosmological and
astrophysical analysis and observations.


Before concluding, it is worth observing that the PQ-mechanism is
not spoiled by the presence of a magnetic field within the limits
allowed by cosmology. This is not obvious since both the magnetic
field and QCD effects break the PQ-symmetry, with the former not
necessarily toward the CP-conserving minimum. If $B$ is large
enough, a temperature $T_*$ would exist such that, for $T \gg
T_*$, the QCD contribution to the axion potential
$\mathcal{L}_{\rm CP} + \mathcal{L}_{ag}$ would be negligible with
respect to the magnetic one.
The Vafa-Witten theorem~\cite{Vafa}, applied to the Lagrangian
with the only electromagnetic contribution,
$\mathcal{L}_{a\gamma}$, leads to the inequality $V[0] \leq V[a]$
for the axion potential energy $V[a]$. Therefore, the axion
dynamically evolves toward that state $a = 0$ which minimizes its
potential.
However, at temperatures sufficiently below $T_*$, the role of
gluons becomes prominent with respect to the electromagnetic
contribution and therefore the term $\mathcal{L}_{a\gamma}$ can be
neglected with respect to $\mathcal{L}_{\rm CP} +
\mathcal{L}_{ag}$. In this case, $V[a]$
satisfies
$V[-\Tb] \leq V[a]$
and so the axion field evolves toward the CP-even minimum of the
Lagrangian, $a = -f_a \Tb$.
The PQ-mechanism is therefore effective for the solution of the
strong CP-problem, unless the magnetic-induced axion mass
dominates the QCD mass until very recent times ($T_* \sim T_{\rm
today}$). This in principle sets a bound on the possible intensity
of a primordial magnetic field. However, it is clear from the
discussion above that, if the magnetic field satisfies the bounds
from BBN and CMB, the magnetic mass $\mb$ is always negligible
with respect to $\mq$ under the QCD-phase transition. Therefore,
we can safely conclude that no magnetic field allowed by the
standard cosmology can spoil the PQ-mechanism.


In conclusion, our analysis shows that a sufficiently intense
cosmological magnetic field could considerably modify the expected
axion relic abundance. For example, assuming the common value $f_a
\simeq 10^{12} \GeV$, we find a threshold value for the magnetic
field at, say, the electroweak scale, $B_{\rm th}(T_{ew}) \simeq 3
\times 10^{21} \G$. A magnetic field more intense than that would
cause a reduction of the expected axion density. This result is
relevant for axion physics and cosmology in general, since axions
are widely believed to be a relevant fraction of the dark matter
in the universe.
In addition, we have shown that a sufficiently intense primordial
magnetic field ($B \gtrsim 10^{24} \G$ at the electroweak scale)
would allow the PQ-constant to be $f_a \sim 10^{15} \GeV$, a scale
that could easily be related to the GUT scale. This would solve
the widely discussed problem of the meaning of the PQ-scale,
requiring no physics beyond the standard model. This result also
shows that the present experiments for the axion search at scales
$f_a \lesssim 10^{12} \GeV$ would not be able to exclude the
existence of the axion field, unless it would also be possible to
prove the non-existence of such an intense primordial magnetic
field.

\vspace*{0.3cm}

\begin{acknowledgments}
We would like to thank G. G. Raffelt for helpful discussions.
\end{acknowledgments}


\end{document}